\UseRawInputEncoding

\documentclass[pdflatex,sn-mathphys-num]{sn-jnl}

\usepackage{graphicx}%
\usepackage{multirow}%
\usepackage{amsmath,amssymb,amsfonts}%
\usepackage{amsthm}%
\usepackage{mathrsfs}%
\usepackage[title]{appendix}%
\usepackage{xcolor}%
\usepackage{textcomp}%
\usepackage{manyfoot}%
\usepackage{booktabs}%
\usepackage{algorithm}%
\usepackage{algorithmicx}%
\usepackage{algpseudocode}%
\usepackage{listings}%

\theoremstyle{thmstyleone}%

\theoremstyle{thmstyletwo}%

\theoremstyle{thmstylethree}%

\begin{document}

\title[Article Title]{ Alternative Representation of the Magnus Series via an Exact Proper Operator Exponen}

\author {\fnm{Yurii \,N.} \sur{Kosovtsov}}\email{yunkosovtsov@gmail.com}

\affil{\city{Lviv}, \country{Ukraine}}

\abstract {This report emphasizes an alternative representation of the Magnus series using proper operator (matrix) exponential solutions for systems of differential equations, including both linear and nonlinear ODEs and PDEs. The main idea is the \emph{exact} \emph{linear} representation of \emph{nonlinear} DEs. Starting from Dyson's time-ordered solutions, we directly convert them into simple proper operator exponents using only generalizations of the well-known Baker-Campbell-Hausdorff (BCH) and Zassenhaus formulae for $t$-dependent operators. This method is explicit both in its operator form and in expressing the formal solution as an ordinary exponential, which simplifies the calculation of analytical solutions as a Taylor series in the variable $t$. By introducing a mutually invertible change of variable for $t$ into the original equations and then solving the new equation with an ordinary exponential, one can obtain a completely different Taylor expansion for the desired function. The essence of this method is the resummation of the series.}


\maketitle

\section{Introduction}
Series expansions of solutions to differential equations have been widely investigated in the literature. In 1949, F. J. Dyson \cite{Dyson1} published a paper in which he introduced the concept of the time-ordered (or chronological) exponent, which made it possible to concisely represent the solution of a homogeneous linear differential equation with varying coefficients, analogous to how an ordinary exponential represents the solution for a linear differential equation with constant coefficients.

For linear ordinary differential equations with non-constant coefficients (also known as non-autonomous ODEs),
\begin{equation}
u' =A(t){u}
\label{dE}
\end{equation}
with the initial condition
\begin{equation}
u|_{t=0} = u_0,
\notag
\end{equation}
where $A(t)$ is a \emph{linear operator} that generally depends on $t$ but does not explicitly depend on the operator $\frac{\partial}{\partial t}$. Under the assumption that $A$ commutes with itself at all times, i.e., $[A(t_1), A(t_2)] = 0$ for all $t_1, t_2$, the solution for $u$ is expressed by the proper exponential of $A$ as $u(t) =\exp \{\int_0^t d\tau\,A(\tau)\}$. In the general case, however, the evolution operator has no known explicit form in terms of $A$ and is instead expressed as
\begin{equation}
u(t) ={\bf T}\exp \{\int_0^t d\tau\,A(\tau)\} u_0,
\label{ET}
\end{equation}
where ${\bf T}$ is the so-called time-ordering operator. The action of this operator transforms the series representation of the ordinary operator-exponential function into an operator-valued Peano-Baker series (also known as the Dyson series \cite{Dyson2}), which involves an infinite sum of nested integrals. This series is unwieldy, hinders theoretical analysis, is computationally expensive, and is purely perturbative from a physical standpoint.

Over time, it became clear that solutions to evolution equations for both nonlinear ODEs \cite{Agrachev,Kosovtsov1,Kosovtsov2} and nonlinear PDEs \cite{Kosovtsov3} and others, as well as for their coupled systems, could also be represented using time-ordered exponents. The main idea here is the \emph{exact} \emph{linear} representation of \emph{nonlinear} DEs. While it is well known that nonlinear ODEs can be represented as linear partial differential equations \cite{Kosovtsov1,Kosovtsov2}, nonlinear PDEs can be \emph{exactly} represented by \emph{linear} differential equations with functional derivatives \cite{Kosovtsov3,Kosovtsov4}.

Almost immediately after the appearance of Dyson's paper \cite{Dyson1}, the question arose of whether it was possible to \emph{exactly} represent time-ordered exponents in a simpler form, without the time-ordering operation.

Magnus's proposal \cite{Magnus1} with respect to the linear evolution equation
\[Y'(t) = A(t)Y (t)\]
with the initial condition $Y(0) = I$, was to express the solution as the exponential of a certain operator $\Omega(t)$,
\begin{equation}
Y (t) = e^{\Omega(t)}.
\label{Mag}
\end{equation}

This is in contrast to the representation (\ref{ET}) in terms of the time-ordering operator introduced by Dyson. While the time-ordering operator ${\bf T}$ is eliminated, the main disadvantage of the Dyson series—its unwieldiness—is significantly increased.

Over the past 50 years since the appearance of Magnus's article, a significant number of works have been published, dedicated to both the development of the Magnus expansion itself and its various applications in many fields of science and technology \cite{Wilcox, Mielnik, Strichartz, Fomenko, Monaco, Blanes, Bauer, Giscard}.

\section{Formal Solutions to First-Order Nonlinear ODEs and PDEs by Time-Ordered Operator Exponents}

Since the representation of solutions for linear ODEs and PDEs in the form (\ref{ET}) is well known, here we present a summary of the results associated with the representation of nonlinear ODEs and PDEs \cite{Kosovtsov3}, for which linear problems are a special case. For clarity, we will limit our discussion to single first-order DEs in $t$. Generalizations to DEs of higher orders in $t$ and systems of coupled equations are straightforward (see \cite{Kosovtsov2,Kosovtsov3}).

The starting point for solving differential equations by the operator method is the \emph{linear} first-order differential equation for the operator ${\bf E}={\bf E}(t)$:
\begin{equation}
\frac{\partial {\bf E}}{\partial t} ={\bf L}(t){\bf E}
\label{dE1}
\end{equation}
with the initial condition
\begin{equation}
{\bf E}|_{t=a} = {\bf I},
\notag
\end{equation}
where ${\bf L}(t)$ is a \emph{linear operator} that generally depends on $t$ but not explicitly on the operator $\frac{\partial}{\partial t}$, and ${\bf I}$ is the identity operator.
The solution to this equation for $t>a$ was obtained by F. Dyson in the form of a time-ordered (or chronological) exponent \cite{Dyson1}
\begin{equation}
{\bf E}(t) ={\bf T}\exp \{\int_a^t d\tau\,{\bf L}(\tau)\},
\notag
\end{equation}
where the exponent is understood to represent a power series expansion with subsequent \emph{chronologization} according to the rule:
\begin{align}
{\bf T}\, \{{\bf L}(\tau_1)\,\ {\bf L}(\tau_2)\dots {\bf L}(\tau_n) \}=&{\bf L}(\tau_{\alpha_1})\,\ {\bf L}(\tau_{\alpha_2})\dots {\bf L}(\tau_{\alpha_n}), \notag\\ &
\tau_{\alpha_1}\geq \tau_{\alpha_2}\geq \dots \geq \tau_{\alpha_n}\notag
\end{align}

The equation adjoint to (\ref{dE1}) for the operator ${\bf E}^{-1}={\bf E}^{-1}(t)$ plays a very important role (see \cite{Agrachev,Kosovtsov3}):
\begin{equation}
\frac{\partial {\bf E}^{-1}}{\partial t} =-{\bf E}^{-1}{\bf L}(t)
\notag
\end{equation}
with the initial condition
\begin{equation}
{\bf E}^{-1}|_{t=a} = {\bf I}. \notag
\end{equation}
It is easy to see, using Dyson's idea, that its solution is
\begin{equation}
{\bf E}^{-1}(t)={\bf T}_0\exp \{-\int_a^t d\tau\,{\bf L}(\tau)\},
\notag
\end{equation}
where the operator ${\bf T}_0$ is the operator of anti-chronological ordering, which acts as follows:
\begin{align}
{\bf T}_0\, \{{\bf L}(\tau_1)\,\ {\bf L}(\tau_2)\dots {\bf L}(\tau_n) \}={\bf L}&(\tau_{\alpha_1})\,\ {\bf L}(\tau_{\alpha_2})\dots {\bf L}(\tau_{\alpha_n}), \notag\\ & \tau_{\alpha_1}\leq \tau_{\alpha_2}\leq \dots \leq \tau_{\alpha_n}\notag
\end{align}

The operator ${\bf E}^{-1}(t)$ is the inverse of ${\bf E}(t)$, i.e.,
\[
{\bf T}_0\exp \{-\int_a^t d\tau\,{\bf L}(\tau)\}\,{\bf T}\exp \{\int_a^t d\tau\,{\bf L}(\tau)\}={\bf I}\,.
\]

For $t<a$, the ordering operators ${\bf T}$ and ${\bf T}_0$ exchange roles. In linear problems, the operators ${\bf T}$ and ${\bf T}_0$ can be seen as describing forward and backward scattering, respectively. Henceforth, it will be assumed that $t>a$.

Time-ordered (chronological) operators must be introduced because the operators in the integrands with different values of the variable $\tau$ may not commute, which causes certain difficulties in manipulating chronological exponents. Nevertheless, an examination of the algebraic properties of chronological exponents allows us to obtain an extensive family of operator identities \cite{Agrachev,Kosovtsov1,Kosovtsov2} that provide the ability to transform and analyze chronological expressions.

The operator ${\bf \Delta}$ is a \emph{derivation} if
\[{\bf \Delta}({\bf A + B })f = {\bf \Delta A} f + {\bf \Delta B} f,\]
\[{\bf \Delta}{\bf A}f = ({\bf \Delta A}) f + {\bf A}{\bf \Delta} f\]
for any differentiable function $f$ and any differentiable linear operators ${\bf A}$ and ${\bf B}$. In other words, a linear operation is a derivation if it satisfies the Leibniz rule. As we will see later, chronological operators with derivations in the exponent, which we will denote for short as
\[{\bf E} ={\bf T}\exp \{\int_a^t d\tau\,{\bf \Delta}(\tau)\}, \]
play an extremely important role here.

Let us consider particular cases of derivation operators ${\bf \Delta}$ and the differential equations that correspond to them.

The (general) solution to the \emph{first-order} nonlinear ODE for $u$ (see \cite{Agrachev,Kosovtsov1})
\begin{equation}
\frac{\partial u}{\partial t} =F(t,u), \qquad u|_{t=a} = c,
\label{ode}
\end{equation}
where $c$ is an arbitrary constant, in the form of a time-ordered exponent is:
\begin{equation}
u(t,c) ={\bf T}_0\exp \{\int_a^t d\tau\,F(\tau,c)\frac{\partial}{\partial c}\}\,c.
\label{odesol}
\end{equation}
Here the \emph{linear} derivation operator is
\begin{equation}
{\bf \Delta_1}(\tau)=F(\tau,c)\frac{\partial}{\partial c}.
\notag
\end{equation}

Note that for nonlinear DEs, this operator differs markedly from the operator $A(t)$ in the linear ODE (\ref{dE}) and its solution (\ref{ET}). This feature is characteristic of all nonlinear DEs, since this result is based on the ability to \emph{exactly} represent a \emph{nonlinear} ODE as a \emph{linear} PDE. The equivalent representation of the nonlinear ODE (\ref{ode}) is the following linear PDE:
\begin{equation}
\frac{\partial u(t,c)}{\partial t} +F(t,c)\frac{\partial u(t,c)}{\partial c}=0, \qquad u(t,c)|_{t=a} = c.
\notag
\end{equation}

There are other interpretations of this exact linearization \cite{Agrachev}, \cite{Giscard} - \cite{ Kosovtsov2}, leading to the same result.

For a nonlinear \emph{partial} differential equation for $u(t,x)$ of the \emph{first order} in the variable $t$ and of arbitrary order in the space variables
\begin{equation}
\frac{\partial u}{\partial t} =F(t,x,u,\dots,D_x^{\alpha_j} u),\qquad u|_{t=a} = c(x),
\label{pde2}
\end{equation}
where $\dots,D_x^{\alpha_j} u$ represents a given finite sequence of derivatives of the function $u(t,x)$ with respect to the space variables $x= (x_1,\dots,x_m)$ and $c(x)$ is an arbitrary function, the (general) time-ordered solution is as follows \cite{Kosovtsov3,Kosovtsov4}:
\begin{equation}
u(t,x) ={\bf T}_0\exp \{\int_a^t d\tau\, \int_{\mathbb{R}^m} d^m \zeta \,\,F(\tau,\zeta,c(\zeta),\dots,D_\zeta^{\alpha_j} c(\zeta))\frac{\delta}{\delta c(\zeta)} \}\,c(x)\,,
\notag
\end{equation}
where $\frac{\delta}{\delta c(\zeta)}$ is the functional derivative. Here, the functional derivative is understood as the linear mapping with the following property:
\[\frac{\delta }{\delta c(\zeta)}c(x)=\delta(x-\zeta),\]
where $\delta(x-\zeta)$ is the Dirac delta function.
The chain rule is also valid in this context.

In this case, the \emph{linear} derivation operator is
\begin{equation}
{\bf \Delta_2}(\tau)=\int_{\mathbb{R}^m} d^m\zeta \,\,F(\tau,\zeta,c(\zeta),\ldots, D_x^{\alpha_j}c(\zeta))\frac{\delta}{\delta c(\zeta)}, \qquad \zeta \in \mathbb{R}^m
\notag
\end{equation}
and the equivalent representation of the nonlinear PDE (\ref{pde2}) is the following linear DE with a functional derivative:
\begin{equation}
\frac{\partial u(t,x,c(x))}{\partial t} +{\bf \Delta_2}(t) u(t,x,c(x))=0, \qquad u(t,x,c(x))|_{t=a} = c(x).
\notag
\end{equation}

\section{Exact Representation of a Time-Ordered Exponent by a Proper Operator Exponent}
Here we need \emph{generalizations} of the well-known Baker-Campbell-Hausdorff (BCH) and Zassenhaus formulae for \emph{$t$-dependent} operators. The classical BCH formula merges the product of two exponential operators into a single one, while the Zassenhaus formula splits an exponential operator into a product of exponential operators \cite{Kosovtsov1,Kosovtsov2}.
\begin{align}
&{\bf T}\exp \{\int_a^t d\tau\,{\bf B}(\tau)\} \,{\bf T}\exp
\{\int_a^t d\tau\,{\bf A}(\tau)\}=\notag \\ &{\bf T}\exp \{\int_a^t
d\tau\,[{\bf B}(\tau)+{\bf T}\exp \{\int_a^\tau d\xi\,{\bf
B}(\xi)\}\,{\bf A}(\tau)\,{\bf T}_0\exp \{-\int_a^\tau d\xi\,{\bf
B}(\xi)\}] \}
\label{BprA}
\end{align}
and
\begin{align}
{\bf T}\exp &\{\int_a^t d\tau\,[{\bf B}(\tau)+{\bf C}(\tau)]\}
={\bf T}\exp \{\int_a^t d\tau\,{\bf B}(\tau)\} \times \notag \\ &
{\bf T}\exp \{\int_a^t d\tau\,{\bf T}_0\exp \{-\int_a^\tau
d\xi\,{\bf B}(\xi)\}\,{\bf C}(\tau)\,{\bf T}\exp \{\int_a^\tau
d\xi\,{\bf B}(\xi)\} \}.
\label{BsumA}
\end{align}
By taking the inverse of both sides of (\ref{BprA}) and (\ref{BsumA}), we can find similar identities for the anti-chronological operators \cite{Kosovtsov2}.
\begin{align}
&{\bf T}_0\exp \{\int_a^t d\tau\,{\bf B}(\tau)\} \,{\bf T}_0\exp
\{\int_a^t d\tau\,{\bf A}(\tau)\}=\notag \\ &{\bf T}_0\exp
\{\int_a^t d\tau\,[{\bf B}(\tau)+{\bf T}\exp \{-\int_a^\tau
d\xi\,{\bf B}(\xi)\}{\bf A}(\tau){\bf T}_0\exp \{\int_a^\tau
d\xi\,{\bf B}(\xi)\}] \}. \notag
\end{align}

The analog of (\ref{BsumA}) is
\begin{align}
&{\bf T}_0\exp \{\int_a^t d\tau\,[{\bf B}(\tau)+{\bf C}(\tau)]\} =
\notag \\ & {\bf T}_0\exp \{\int_a^t d\tau\,{\bf T}_0\exp
\{\int_a^\tau d\xi\,{\bf B}(\xi)\}{\bf C}(\tau){\bf T}\exp
\{-\int_a^\tau d\xi\,{\bf B}(\xi)\} \} \times \notag \\ & \qquad \qquad \qquad {\bf
T}_0\exp \{\int_a^t d\tau\,{\bf B}(\tau)\}. \label{BsumA0}
\end{align}
We will also need the well-known shift operator property
\begin{equation}
\exp\{\alpha \frac{\partial}{\partial s}\}\Phi(s)=\Phi(s+\alpha).
\label{shift}
\end{equation}

Now, let us consider the following. Consider the following operator chain:
\begin{align}
&\exp\{(t-a)\,[{\bf \Delta}(s)+\frac{\partial}{\partial s}]\}\overset{(\ref{BsumA0})}{=}\notag \\
& {\bf T}_0\exp\{\int_a^t
d\tau\,\exp\{(\tau-a)\frac{\partial}{\partial
s}\}{\bf \Delta}(s)\exp\{(a-\tau)\frac{\partial}{\partial
s}\}\}\,\exp\{(t-a)\frac{\partial}{\partial
s}\}\overset{(\ref{shift})}{=}\notag
\\ & {\bf
T}_0\exp\{\int_a^t d\tau\,{\bf \Delta}(s+\tau-a)\}\,\exp\{(t-a)\frac{\partial}{\partial s}\}. \notag
\end{align}

Therefore,
\begin{align}
& {\bf
T}_0\exp\{\int_a^t d\tau\,{\bf \Delta_1}(s+\tau-a)\}\,\exp\{(t-a)\frac{\partial}{\partial s}\}c= \notag\\
&{\bf T}_0\exp\{\int_a^t
d\tau\,{\bf \Delta_1}(s+\tau-a)\}\,c
=\exp\{(t-a)\,[F(s,c)\,\frac{\partial}{\partial
c}+\frac{\partial}{\partial s}]\}\, c\notag
\end{align}
and since it is clear that
\begin{align}
& {\bf
T}_0\exp\{\int_a^t d\tau\,{\bf \Delta_1}(s+\tau-a)\}c \,|_{s=a}={\bf T}_0\exp\{\int_a^t d\tau\,{\bf \Delta_1}(\tau)\}c\overset{(\ref{odesol})}{=} \notag\\
&u(t,c)=\exp\{(t-a)\,[F(s,c)\,\frac{\partial}{\partial c}+\frac{\partial}{\partial s}]\}\, c\, |_{s=a}\,,\notag
\end{align}
Thus, we have obtained the solution to ODE (\ref{ode}) in the form of a \emph{proper} operator exponent:
\begin{equation}
u(t,c)=\exp\{(t-a)\,[F(s,c)\,\frac{\partial}{\partial c}+\frac{\partial}{\partial s}]\}\, c\, |_{s=a}\,.
\notag
\end{equation}

Denoting
\begin{equation}
U(t,s,c) =\exp\{(t-a)\,[F(s,c)\,\frac{\partial}{\partial
c}+\frac{\partial}{\partial s}]\}\, c\,,\notag
\end{equation}
we obtain the (general) solution of ODE (\ref{ode}) from the preceding expressions as
\begin{equation}
u(t,c) =U(t,s,c)\,|_{s=a}\,,\notag
\end{equation}
where the function $U(t,s,c)$ satisfies the following \emph{linear} PDE:
\begin{equation}
\frac{\partial U(t,s,c)}{\partial t}-F(s,c)\,\frac{\partial
U(t,s,c)}{\partial c}-\frac{\partial U(t,s,c)}{\partial
s}=0\,,\qquad U(t,s,c)\,|_{t=a}=c. \notag
\end{equation}

Similarly, for the nonlinear PDE (\ref{pde2}), we start from:
\begin{align}
{\bf
T}_0 & \exp\{\int_a^t d\tau\,{\bf \Delta_2}(s+\tau-a)\}\,\exp\{(t-a)\frac{\partial}{\partial s}\}c(x)= \notag\\
&{\bf T}_0\exp\{\int_a^t
d\tau\,{\bf \Delta_2}(s+\tau-a)\}\,c(x)=\notag\\
&\exp\{(t-a)\,\int_{\mathbb{R}^m} d^m\zeta \,\,[F(s,\zeta,c(\zeta),D_x^{\alpha_j}c(\zeta))\frac{\delta}{\delta c(\zeta)}+\frac{\partial}{\partial s}]\}\, c(x)\notag
\end{align}
we get the (general) solution in the form of a \emph{proper} operator exponent:
\begin{equation}
u(t,c(x)) =[\exp\{(t-a)\,\int_{\mathbb{R}^m} d^m\zeta \,\,[F(s,\zeta,c(\zeta),D_x^{\alpha_j}c(\zeta))\frac{\delta}{\delta c(\zeta)}+\frac{\partial}{\partial s}]\}\, c(x)]|_{s=a}.\label{pdesol}
\end{equation}

Now, denoting
\begin{equation}
V(t,c(x),s) =\exp\{(t-a)\,\int_{\mathbb{R}^m} d^m\zeta \,\,[F(s,\zeta,c(\zeta),D_x^{\alpha_j}c(\zeta))\frac{\delta}{\delta c(\zeta)}+\frac{\partial}{\partial s}]\}\, c(x)\,,
\notag
\end{equation}
we obtain the (general) solution of PDE (\ref{pde2}) from the preceding expression as
\begin{equation}
u(t,c(x)) =V(t,c(x),s)\,|_{s=a}\,,\notag
\end{equation}
where the function $V(t,c(x),s)$ satisfies the following \emph{linear} equation:
\begin{equation}
\frac{\partial}{\partial t} V(t,c(x),s)-{\bf \Delta_2}(s)V(t,c(x),s)-\frac{\partial}{\partial
s}V(t,c(x),s)=0,\qquad V(t,c(x),s)\,|_{t=a}=c(x). \notag
\end{equation}

These final forms of the operator solutions have some interesting features—namely, they do not contain the ordering operator ${\bf T}$ or any integrals.

Thus, we have solved the problem posed by Magnus, which can be formulated in two ways: how to eliminate the time-ordering operators from Dyson's solutions to DEs, or, alternatively, how to "correct" the proper operator exponent so that it yields an exact solution in the simplest \emph{explicit} form.

Magnus, to solve this problem, started from the \emph{nonlinear} equation for $\Omega(t)$ in (\ref{Mag})—a difficult task in itself—and obtained an \emph{infinite series} known as the Magnus expansion. In contrast, we start directly from Dyson's time-ordered solutions and, using only \emph{generalizations} of the BCH and Zassenhaus formulae for \emph{$t$-dependent} operators, convert them directly into simple proper operator exponents.

Generalizations for differential equations of any order in $t$ and for coupled systems of such equations are quite straightforward \cite{Kosovtsov2,Kosovtsov3}.

Compared to the cumbersome expressions in the Magnus series method, our approach allows \emph{approximate} solutions for $u$ (from (\ref{odesol}) and (\ref{pdesol})) to be immediately expressed as formal Taylor series in $t$ without solving auxiliary systems of equations. The \emph{Maple} procedures for analytical calculations of approximate solutions to linear and nonlinear ordinary and partial differential equations (Cauchy problem), which are based on the approach considered here, are presented in \cite{Kosovtsov5}.

In addition, the method described above, being explicit both in terms of the operator and in terms of expressing the formal solution as an ordinary exponential, makes it quite easy to calculate analytical solutions in the form of a Taylor series in the single variable $t$. Generally, the resulting Taylor series may not converge quickly, and this version often lacks a small parameter. We will not address the convergence of the formal series obtained here (see \cite{Kosovtsov4}), but we will briefly discuss a related extension.

\section{Options for Representing Solutions as Generalized Taylor (Fourier) Series}

If we introduce a mutually invertible change of the variable $t$ into the original equations, find a solution to the new equation (either as an ordinary exponential or as a Taylor series using the method described here), and then revert to the original variable $t$, we obtain a completely different Taylor expansion of the sought function. For example, using the substitution $\tau=\exp(-pt)$, where $p$ is an arbitrary parameter (real or complex), we obtain a series in terms of $\tau=\exp(-pt)$. If the real part of $p$ is positive, this can ensure faster convergence. In this case, $1/(\mathfrak{Re} \,p)$ acts as an artificially introduced small parameter that allows for control of the remainder term. By analogy, one can use many other mutually invertible substitutions to achieve different rates of convergence. The essence of this method is the resummation of the series.

This approach is illustrated with the following nontrivial PDE:
\begin{equation}
\frac{\partial^2u(t, x)}{\partial t^2} = u(t, x)\frac{\partial^2 u(t, x)}{\partial t\,\partial x} ,\qquad u(0,x)=h(x),\,\frac{\partial u(t, x)}{\partial t}\,|_{t=0}=g(x).
\label{pde3}
\end{equation}
Expanding the operator exponent (\ref{pdesol}) corresponding to equation (\ref{pde3}) into a series and applying the operators, we obtain a solution to the PDE in the form of a Taylor series (to avoid unwieldy expressions, we show only the first five terms):
\begin{align}
u(t,& x) = h + g t + h g'\,\frac{t^2}{2} +(h^2 g'' + g'(h h' + g))\,\frac{t^3}{6}+\notag\\&
(h h'^2g'+3 h^2 h' g''+h^2 h'' g'+h^3 g''' +2 h' g g'+2 h g'^2+3 h g g'' )\frac{t^4}{24}
+\ldots
\label{anspde3}
\end{align}

Now, let us apply the series resummation method to the same equation (\ref{pde3}).
By making the mutually invertible change of variable $t = -\ln(\tau)/p$ (or $\tau = e^{-p\,t}$) in equation (\ref{pde3}), we obtain the following \emph{auxiliary} nonlinear equation:

\begin{equation}
W(\tau, x)\frac{\partial^2 W(\tau, x)}{\partial\tau\,\partial x}+(\tau\,\frac{\partial^2 W(\tau, x)}{\partial\tau^2} + \frac{\partial W(\tau, x)}{\partial \tau})\,p=0
\notag
\end{equation}
with new initial conditions that are consistent with those of the original PDE:
\[W(1,x)=h(x),\,\,\frac{\partial W(\tau, x)}{\partial \tau}\,|_{\tau=1}=-g(x)/p.\]

Using the same method as for the original PDE, we obtain a solution to the auxiliary PDE as a Taylor series in $\tau$. Finally, we apply the inverse substitution $\tau = e^{-p\,t}$ to this result. As a result, we obtain a new solution to PDE (\ref{pde3}) in the form of a generalized Taylor series:
\begin{align}
u&(t, x) =h- \frac{g}{p}(e^{- p\,t}-1)+\frac{1}{2 p^2}(hg'+gp)(e^{- p\,t}-1)^2 +\notag\\
& \frac{1}{6 p^3}[-h^2 g''+(-g-3hp-hh')g'-2gp^2](e^{- p\,t}-1)^3+\notag\\
&\frac{1}{24 p^4}[h^3g'''+(3hg+6h^2p+3h^2h')g''+h^2h''g'+2hg'^2+\notag\\
&(hh'^2+(6hp+2g)h'+11hp^2+6gp)g'+6gp^3](e^{- p\,t}-1)^4+\ldots
\label{anspde4}
\end{align}

It can be shown that for \emph{any} $p$, the expansion of this last series (\ref{anspde4}) into an ordinary Taylor series is identical to the series in (\ref{anspde3}).

If we set $p=i\omega$ in (\ref{anspde4}), where $\omega \in \mathbb{R}$, we obtain the solution to PDE (\ref{pde3}) as a \emph{Fourier series}. Typically, Fourier expansion is used for linear DEs with coefficients independent of $t$. The proposed method can be directly applied to obtain solutions as Fourier series for both linear and nonlinear DEs with $t$-dependent parameters.

\end{document}